# Systematic discovery of new nano-scale metastable intermetallic eutectic phases in laser rapid solidified Aluminum-Germanium alloy

Arkajit Ghosh [a, *], Wenqian Wu [b], Tao Ma [c], Ashwin J. Shahani [a], Jian Wang [b], Amit Misra [a, d, *]

[a] Department of Materials Science and Engineering, University of Michigan – Ann Arbor, MI 48109, USA
[b] Department of Mechanical and Materials Engineering, University of Nebraska – Lincoln, NE 68588, USA
[c] Michigan Center for Materials Characterization, University of Michigan – Ann Arbor, MI 48109, USA
[d] Department of Mechanical Engineering, University of Michigan – Ann Arbor, MI 48109, USA

*Corresponding authors' email addresses: arkajitg@umich.edu; amitmis@umich.edu

## Abstract

Laser surface remelting of as-cast Al-Ge eutectic alloy is shown to produce ultrafine lamellar eutectic morphology with interlamellar spacing refined up to ~60 nm and composed of FCC Al solid solution and unusual $Al_xGe_y$ intermetallic phases that do not form during near-equilibrium solidification. The microstructures are characterized and analyzed using a combination of selected area electron diffraction, high-resolution scanning transmission electron microscopy, energy dispersive X-ray spectroscopy to obtain high-resolution elemental maps, and atomistic modeling using density functional theory followed by atomic-scale image simulation. Depending on the local solidification conditions, the crystallography of the $Al_xGe_y$ intermetallic phases in the eutectic microstructure is either monoclinic (C 2/c) or monoclinic (P $2_1$), with high densities of defects in both cases. This is in sharp contrast to the as-cast alloys that showed nominally pure Al and Ge phases with significant solute partitioning and equilibrium FCC and diamond cubic crystal structures, respectively. Corresponding kinetic phase diagrams are proposed to interpret the evolution of nano-lamellar eutectic morphologies with equilibrium Al and metastable $Al_xGe_y$ phases, and to explain increased solid solubility in the Al phases manifested by precipitation of ultrafine clusters of Ge. The reasons for the formation of these metastable eutectics under laser rapid solidification are discussed from the perspective of the competitive growth criterion.

**Keywords**: Al-Ge eutectic, laser rapid solidification, new intermetallic phases, metastable phase equilibria, high-resolution scanning transmission electron microscopy, density functional theory.

## Graphical Abstract

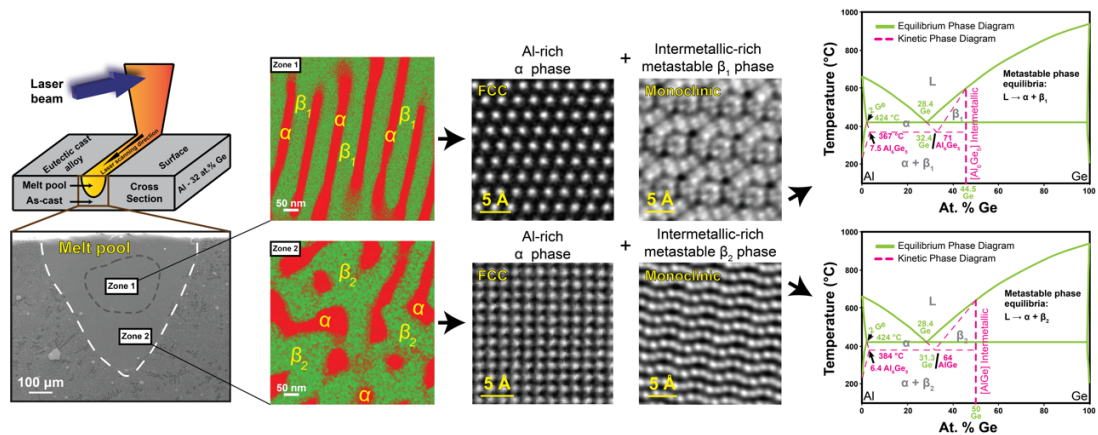

# 1. Introduction

Metallic materials often exhibit metastability during evolution from synthesis to applications, which are either circumstantially induced by processing or purposefully engineered [1, 2, 3]. Synthesis or processing inherited metastability could be broadly classified into two categories: (i) compositional and (ii) structural [4]. An example of the first includes solid state phase transformation, while the latter is usually manifested through formation of intermetallic compounds or amorphous solids or unusual defect structures [1,4]. Conventional processing approaches to incorporate metastability encompass but are not limited to vapor or melt or solid quenching, mechanical cold working or alloying, and irradiation [4, 5, 6]. Rapid solidification, i.e., ultrafast melt quenching, has been performed on different alloy systems to incur both compositional and structural metastability, which produce exotic microstructures with extraordinary physical properties [7, 8, 9, 10]. Although the recent trend of engineering metastable microstructure primarily focuses on multi principal element alloys, bicomponent eutectic alloys are not exceptions in exhibiting metastability [5, 11, 12, 13].

Rapid solidified eutectic alloys comprising soft and hard phases are good model systems to study deformation mechanisms responsible for simultaneous enhancement in strength and ductility in disparate phase ultrafine materials, that have recently garnered significant research interest [14, 15]. Laser rapid solidification, that is reportedly used to impart high cooling rate of up to $10^7$ K/s [16], has been extensively used in order to refine the as-cast micron-scale eutectic microstructures of such eutectics, e.g., Al-Si and Al-Al$_2$Cu alloys, to nano-scale fibrous and lamellar structures, respectively [17, 18, 19, 20]. Metastability in those nano-scale structures are reflected to some extent by (i) precipitation due to increased solid solubility as the solidus and liquidus curves approach $T_0$ lines of the kinetic eutectic phase diagrams during rapid solidification [21], (ii) solute segregation as a result of complete solute trapping ahead of the solid-liquid interface [16], and (iii) solidification instigated defects due to induced shearing [22]. However, large degree of structural transformation (away from the equilibrium prediction), such as formation of metastable intermetallics or amorphization, has not been reported for the foregoing eutectic systems with laser processing. This necessitates design of laser rapid solidification experiments with new alloy candidates, that might bring about a large scale metastable structural transition.

Al-Ge binary system produces faceted fishbone-like lamellar eutectic microstructures with α-Al (FCC) and β-Ge (diamond cubic) phases during equilibrium solidification [23]. Amorphous or metastable crystalline phases have been reported in mechanically alloyed or splat cooled Al-Ge alloys [24, 25, 26], with crystal structures such as hexagonal close packed (HCP), monoclinic, orthorhombic, as determined by reciprocal space imaging via transmission electron microscopy (TEM) or X-ray diffraction (XRD). Reciprocal space imaging for unknown phases comes with certain shortcomings while determining crystal structures particularly in case there are very weak or very strong peaks due to atomic ordering. On the other hand, bulk scale characterization such as XRD may not be feasible for laser processed microstructures that are often confined within a melt pool having depth and diameter of few hundred μm [16, 17]. From both of these perspectives, real space imaging in addition to reciprocal space characterization is necessary to accurately index the metastable phases. Besides, the microstructure evolution is still unknown in Al-Ge system for laser rapid solidification, which induces larger undercooling and higher cooling rate compared to other solidification techniques, e.g., melt spinning, splat quenching, used in previous studies to generate metastable phases in the same alloy [24, 26], resulting in nano-scale eutectics.

This work intends to study the microstructure of laser rapid solidified Al-Ge eutectics with regard to length-scale, morphology, and phase evolution across melt pool. Efforts have been made to fully characterize the metastable phases, produced in solid solution (Al-rich) – intermetallic ($Al_xGe_y$) eutectic reactions, with the aid of energy dispersive X-ray spectroscopy (EDX) and high resolution scanning transmission electron microscopy (HR-STEM). The monoclinic $Al_6Ge_5$ and AlGe intermetallics have been observed for the first time and this work reports the structural details at atomic-level by reciprocal as well as real space imaging and validating them by atomic-scale structural modeling. Composition and crystallography of these structures are significantly different than the prior reported monoclinic phases in Al-Ge systems. The experimentally designed unit cells could be successfully relaxed by density functional theory (DFT), which has also helped refining the structural parameters. The findings of this research reveal that not all the disparate phase eutectics produce the expected equilibrium phases (e.g., Al-Si or $Al-Al_2Cu$), when subjected to laser induced non-equilibrium solidification, and the metastable phase formation may be kinetically favorable by requiring little solute partitioning. . Kinetic phase diagrams corresponding to the metastable $Al-Al_xGe_y$ eutectics have been outlined and analyzed to correlate with the eutectic morphologies and precipitates in the Al-rich phases (an attribute of the compositional metastability). The selection of microstructure has been further discussed by invoking the competitive growth criterion. Apart from the structural metastability displayed through the unusual intermetallic rich phases, another degree of structural metastability has been prevalent by the formation of solidification induced defects such as dislocations and stacking faults within the eutectic phases.

## 2. Methodology

### 2.1. Material and processing

Hyper-eutectic Al – 56 wt. % Ge [Al – 32 at. % Ge] as-cast alloy was prepared by vacuum arc melting. A slightly hyper-eutectic starting composition was planned to compensate the deviation of the eutectic point during non-equilibrium solidification, which is inevitable due to shift in the solidus and liquidus curves toward $T_0$ lines in a state of complete solute trapping at extreme cooling rates as in laser rapid solidification [27]. From the as-received buttons, rectangular pieces with dimensions 20 mm (length: L) × 10 mm (width: W) × 5 mm (thickness: T) were prepared using diamond wafer saw. The top surfaces were ground with 120-grit SiC sandpaper in order to ensure maximum laser absorption and uniform melt pool microstructure throughout the length of the scanning track [16, 28]. Single scan laser surface remelting (LSR) experiments were carried out on the ground surface of the specimen in Open Additive PANDA™ machine with laser power, scanning velocity, and fine spot diameter of 200 W, 100 mm/s, and 75 μm, respectively, as reported earlier for Al-Si system [16].

### 2.2. Microstructure characterization

Scanning electron microscopy (SEM) with TFS Helios 650 NanoLab SEM/FIB was done using backscattered electron (BSE) imaging mode at accelerating voltage and beam current of 5 kV and 0.2 nA, respectively. BSE micrographs exhibit high contrast owing to the large atomic number difference between Al and Ge. Cross-section samples were lifted out and thinned using focused ion beam (FIB) for transmission electron microscopy (TEM) and scanning transmission electron microscopy (STEM). Low magnification STEM and STEM-EDX of up to 510,000X were done with TFS Talos F200X G2 using 200 kV accelerating voltage to characterize morphology, phase distribution, and to build tomography reconstruction. 0.15 nm spot size was used for STEM-EDX experiments. Background subtraction in the elemental maps was done by using a prefiltering with 5 px average filter and postfiltering with a Gaussian blur having sigma 2.0. STEM tomography experiments were conducted by using a Fischione high-tilt high-visibility tomography holder and tilting the stage from -70° to +70° with an interval of 2°. Three dimensional reconstruction was made with the aid of Gatan GMS 3 software. Atomic-scale images and selected area electron diffraction (SAED) patterns were captured with an aberration corrected TFS Spectra (S)TEM at 300 kV accelerating voltage and 75 pA screen current. For atomic resolution EDX mapping, a 0.05 nm spot size was used, screen current was increased to 100 pA, and the postfiltering was selected as Radial Wiener filter with highest frequency of 60. Atomic-scale images were collected in stack and processed by drift corrected frame integration in Velox software. High angle annular dark field (HAADF), integrated differential phase contrast (iDPC), annular bright field (ABF), and dualX EDX detectors were used to index the atomic structures in detail.

### 2.3. Atomistic modeling

VESTA software [29] was used to build the three-dimensional atomic models based on the experimental two-dimensional projected atomic resolution STEM images from multiple zone axes. Crystallographic information files (CIFs) obtained from VESTA were used for density functional theory (DFT) calculations followed by HR-STEM image simulations of the DFT relaxed structures. Vienna Ab initio Simulation Package (VASP) [30] was used for DFT based geometric optimization (relaxation) of the experimentally obtained structures. The projector augmented wave method [31] was utilized to describe the interaction between ions and electrons, with the plane wave cutoffs set at 500 eV. The Perdew–Burke–and Ernzerhof exchange correlation based on the generalized gradient approximation was employed [32]. The Al-s2p1 and Ge-3d4s4p states were treated as valence electrons. Γ-centered k-point grids of 6×6×10 for $\beta_1$ (first intermetallic-rich phase) unit cell and 12×8×6 for $\beta_2$ (second intermetallic-rich phase) unit cell were used for integration over the first Brillion zone during structural optimizations. Convergence criteria were set at $1.0 \times 10^{-5}$ eV for energy and 0.01 eV/ Å for force. Multi-slice image simulations of the DFT-relaxed structures were done for ~20 nm thick specimens from different low index zone axes using abTEM package [33] to compare with the corresponding HR-STEM micrographs.

## 3. Results

### 3.1. Microstructure across the melt pool: Morphology and phase evolution

Figure 1a illustrates the SEM image of the melt pool that was formed after LSR on the sample cross-section. Typical faceted lamellar eutectic microstructure with chunks of primary Ge (feature of hypereutectic as-cast microstructure) can be observed through the BSE micrograph outside the melt pool in figure 1b. There are two distinct zones within the melt pool based on the microstructural features, labelled zone 1 and zone 2, as can be seen from the BSE micrographs in figure 1(c-d). General microstructural features in both of these zones are quite different from their as-cast counterpart, not only in terms of refinement, but also in terms of morphology. Based on the atomic number contrast in the BSE micrographs, a standard nomenclature of the phases will be followed henceforward: The dark and the bright phases in zone 1 and zone 2 will be termed as $\alpha_1$, $\beta_1$, and $\alpha_2$, $\beta_2$, respectively. In zone 1, uniform lamellar or plate-like structure of $\alpha_1$ and $\beta_1$ is prevalent which gets coarser at the cell boundaries. The average interlamellar spacing is ~ 60 nm, with plate thickness ratio ($\alpha_1$:$\beta_1$) of ~ 2:3. Figure 1e presents the perfect plate-like structure from different angles as reconstructed from STEM tomography experiments. In zone 2, the plate like structure is not as continuous as in zone 1. Average interphase spacing is ~ 90 nm with plate thickness ratio ($\alpha_2$:$\beta_2$) of ~ 1:2. It is evident from the STEM tomography reconstruction in figure 1f that the lamellar eutectic morphology is degenerated in this zone. It is noteworthy though that there are local deviations (such as coarsening) in geometry and spacing in both of the zones, where the average values may seem inconsistent.

In order to determine the elemental concentrations within different phases, elemental maps were taken from the representative regions from zone 1 and zone 2. Figure 2 (a1-a4) and figure 2 (c1-c4) display HAADF STEM micrograph, Al partitioned EDX map, Ge partitioned EDX map, and combined elemental EDX maps from zone 1 and zone 2, respectively. Areas denoted by 1, 2, 3, and 4, are representative regions from the phases $\alpha_1$, $\beta_1$, $\alpha_2$, and $\beta_2$, respectively, in figure 2a1 and figure 2c1. Spectra received from areas 1 and 3 (figure 2b1 and figure 2d1) depict that the $\alpha_1$ and $\alpha_2$ (dark) phases are Al-rich and the negligible composition difference between them (within error limit) indicates both of them could be same phases. On the other hand, spectral analyses from the other two areas (figure 2b2 and figure 2d2) indicate that the $\beta_1$ and $\beta_2$ (bright) phases contain both Al and Ge, suggesting possible intermetallic-rich metastable structures. Detailed quantification of elemental concentrations in different areas has been given in Table 1.

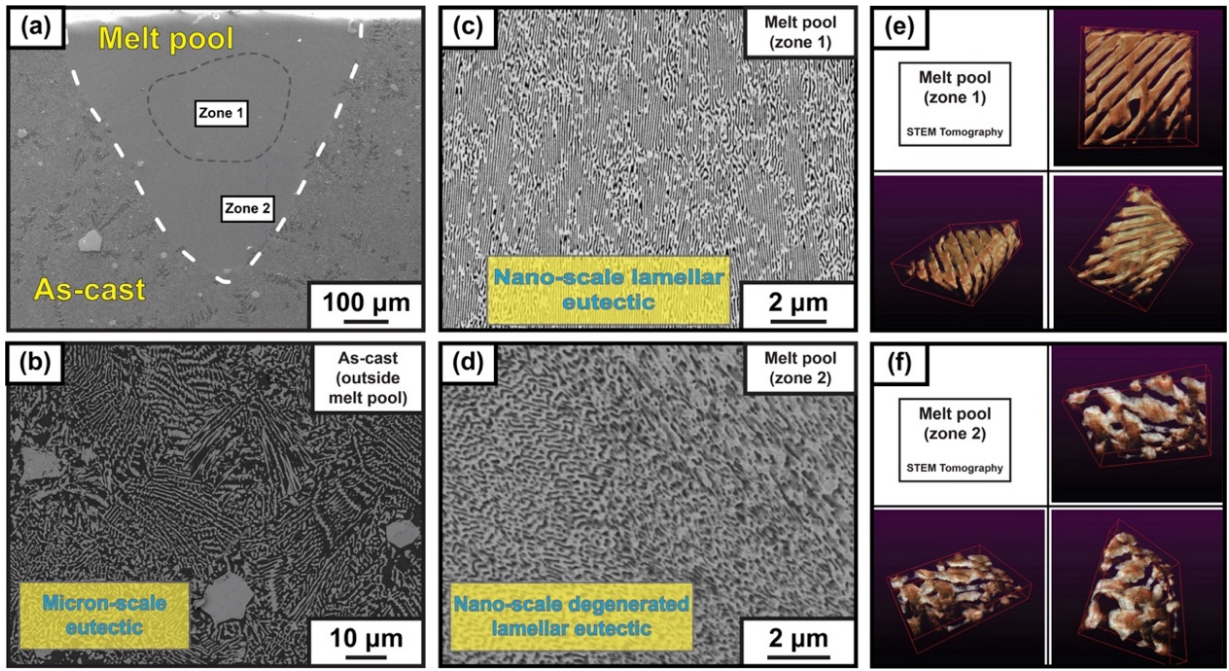

Figure 1: (a) SEM image of the melt pool at the sample cross-section due to laser surface remelting (LSR) having two distinct microstructural zones: zone 1 and zone 2. (b) BSE micrograph of typical as-cast hypereutectic Al-Ge microstructure outside the melt pool. BSE micrographs of the ultrafine two-phase structures within (c) zone 1 and (d) zone 2 of the melt pool. STEM tomography reconstructions reveal that the non-equilibrium phase morphologies in (e) zone 1 is continuous lamellar, whereas in (f) zone 2, the lamellar morphology is degenerated. Supplementary videos S1 and S2 provides with the videographic visualization of the tomography reconstructions.

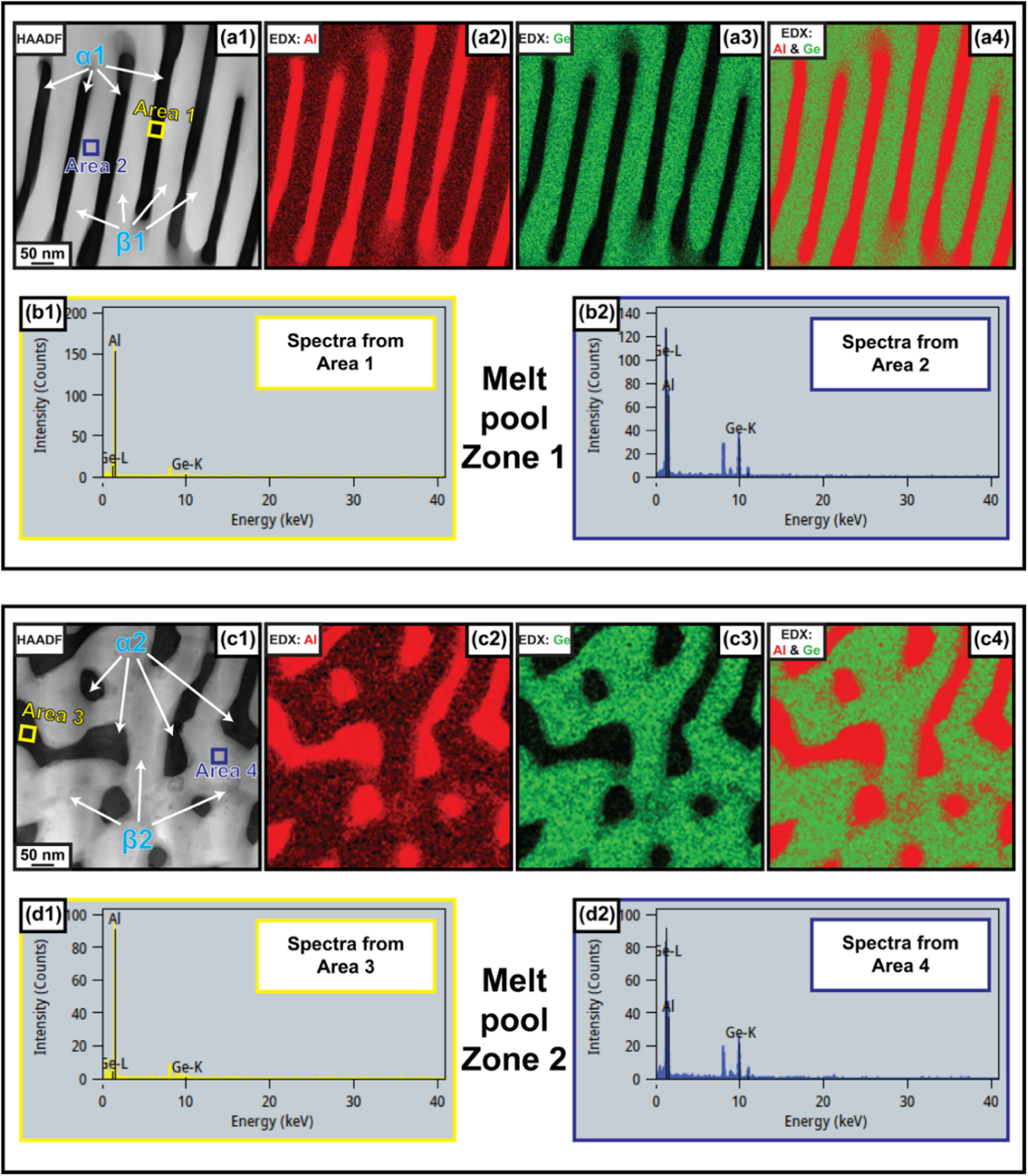

Figure 2: HAADF STEM micrographs, Al partitioned elemental maps, Ge partitioned elemental maps, and mixed elemental maps obtained from STEM-EDX experiments for (a1-a4) zone 1 and (c1-c4) zone 2 within the melt pool. Areas 1, 2, 3, and 4 represent the phases $\alpha_1$, $\beta_1$, $\alpha_2$, and $\beta_2$, respectively. EDX spectra received from (b1) area 1, (b2) area 2, (d1) area 3, and (d2) area 4 indicate that the $\alpha_1$ and $\alpha_2$ phases are Al-rich, whereas the $\beta_1$ and $\beta_2$ phases are Al-Ge intermetallic rich.

Table 1: <u>STEM-EDX obtained elemental quantification of different phases in zone 1 and zone 2</u>

| Area | Representative phase | Al concentration (at.%) | Ge concentration (at.%) | Remarks |
|---|---|---|---|---|
| #1 | $\alpha_1$ | 97.4 $\pm$ 1.2 | 2.6 $\pm$ 1.2 | Al-rich |
| #2 | $\beta_1$ | 54.8 $\pm$ 3.2 | 45.2 $\pm$ 3.2 | Intermetallic-rich |
| #3 | $\alpha_2$ | 97.7 $\pm$ 1.2 | 2.3 $\pm$ 1.2 | Al-rich |
| #4 | $\beta_2$ | 50.5 $\pm$ 3.3 | 49.5 $\pm$ 3.3 | Intermetallic-rich |

It is important to find out the crystallographic details of all these phases, and SAED along with HR-STEM approaches have been extensively used to index the atomic structures throughout next section. It is quite relevant to mention here that the intermetallic-rich phases do not form during the equilibrium solidification conditions; the corresponding as-cast microstructural features have been described in supplementary figure S2.

### 3.2. Crystallography of the phases within melt pool: Al-rich as-equilibrium and intermetallic-rich metastable phases

Figure 3 (a-b) present HAADF STEM micrographs of zone 1 and zone 2, out of which the $\alpha_1$ and $\alpha_2$ phases are chosen for high resolution imaging. HAADF HR-STEM images given in figure 3 c1 and 3 c2 have been indexed and found to be along [0 1 1] and [0 0 1] zone axes of a face centered cubic (FCC) crystal. Spacings between (111) planes [$d_{111}$ = 0.233 nm] calculated from [0 1 1] zone axis image and between (0 0 2) planes [$d_{002}$ = 0.202 nm] calculated from [0 0 1] zone axis match exactly with the crystallographic interplanar spacings of Al, and hence it can be concluded that the dark $\alpha_1$ and $\alpha_2$ phases are governed by Al-rich Al-Ge solid solutions, which is also in line with the elemental quantification from EDX (Table 1). Therefore, both $\alpha_1$ and $\alpha_2$ phases are actually the same Al-rich solid solution phases, and they have been termed as $\alpha$ henceforth. Since the elemental concentration in the possible metastable phases are different in two zones, indexing of $\beta_1$ and $\beta_2$ phases are being reported separately unlike the $\alpha$ phases.

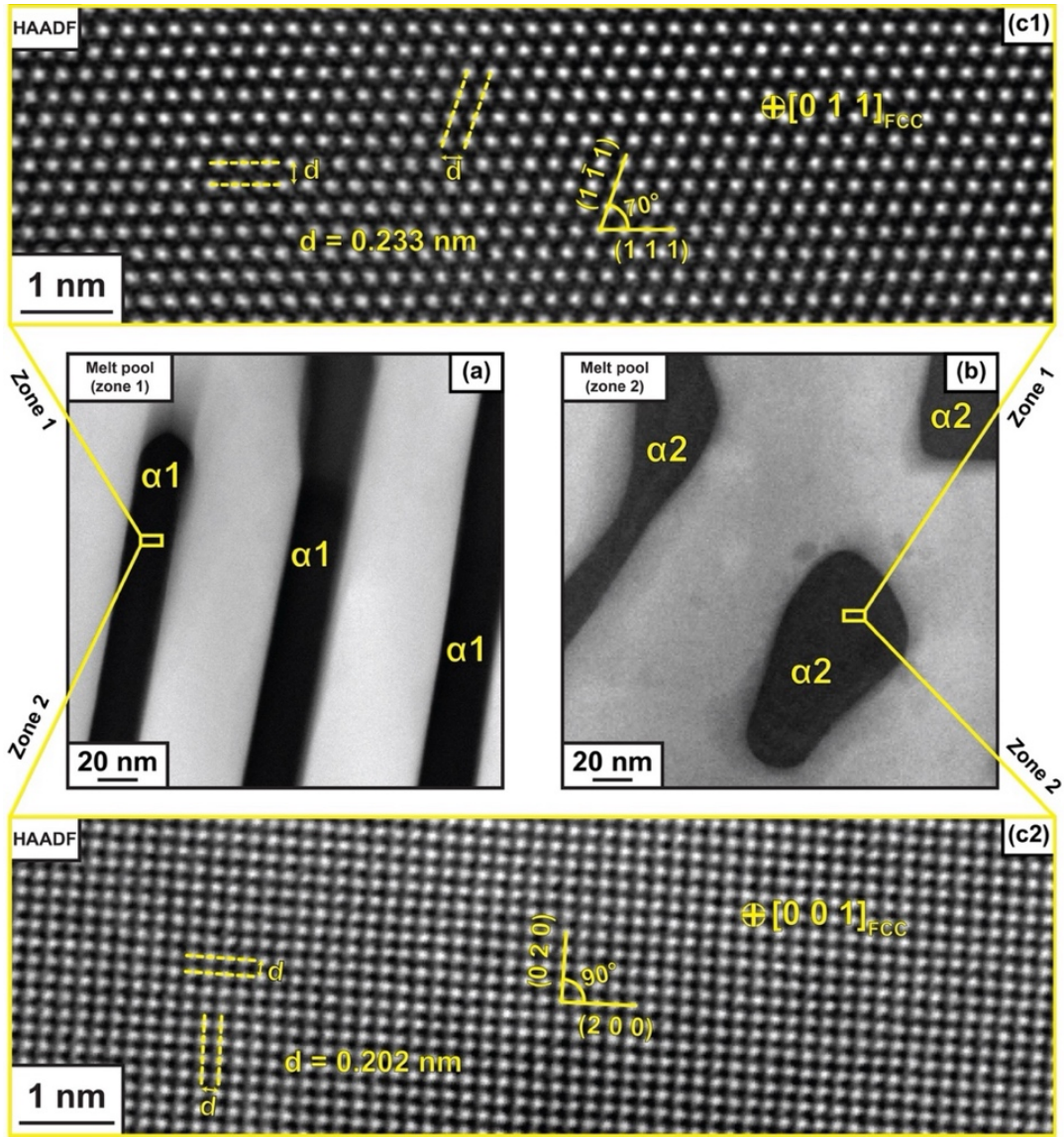

Figure 3: HAADF STEM images from (a) zone 1 and (b) zone 2 with the $\alpha_1$ and $\alpha_2$ phases highlighted. HAADF HR-STEM images from (c1) [0 1 1] and (c2) [0 0 1] zone axes have been indexed as FCC Al crystal. Therefore, the $\alpha_1$ and $\alpha_2$ phases can be considered similar to the equilibrium $\alpha$ eutectic phase in Al-Ge system.

Figure 4a maps series of SAED patterns taken from different zone axes of $\beta_1$ and mapped according to their relative angular positions. Calculation of the relative angles based on the tilting angles for a double tilt TEM holder can be found elsewhere [34]. Three different low index zone axes were marked across a Kikuchi band (colored red) and respective low index planar vector has been denoted as g1. Planar peaks perpendicular to g1 were plotted according to their relative angular difference to reconstruct the two dimensional unit lattice orthogonal to g1 in figure 4b. Possible missing planes have been marked by 'x'. Since the angular range of the SAEDs plotted across the Kikuchi band is as much as ~125° by collecting the diffraction patterns consistently from a few adjacent lamellae (consistency was ensured by measuring relative angles between the zone axes), it has been possible to image two principal reciprocal planes (a'-b' and c'-b') by measuring other low index reciprocal planar vectors g2 and g6, and also predict the angle in between them. Hence, all the reciprocal lattice parameters (a', b', c', α', β', and γ') could be determined and inverted to visualize the real space lattice, with parameters as listed in table 2 that deduce a monoclinic Bravais lattice. The entire procedure of reciprocal cell reduction used here is according to Niggli reduction model used for indexing unknown crystals [35]. The manual calculation by wide range tilting reduced the mathematical complexity of using a Niggli matrix. The mode of distribution of the planar peaks (analysis of the missing peaks for specific zone axes) and their intensities in low index (LI) zone convergent beam electron diffraction (CBED) patterns indicate a possible C 2/c space group. Monoclinic (C 2/c) lattice having the derived parameters implies a new phase in the Al-Ge system, for which estimating the atomic positions (motifs) is equally important. HR-STEM images in figure 5 simultaneously captured at different imaging conditions like HAADF and iDPC from three low indexed zones ([001], [101], [100]) along with atomic resolution EDX from one zone having largest average spacing between the atomic columns ([101] here) have enabled near-precise (measurement precision is 0.05) estimation of the atomic positional coordinates, given in table 3. HAADF-STEM images are useful to locate the position of Ge atoms because of the bright contrast that arises from higher atomic number. Simultaneous iDPC-STEM images point out the positions of the low atomic number element Al, the signal of which are much weaker in HAADF-STEM. Atomic scale EDX could reveal the nature of the atomic columns, e.g., the brighter and darker signals in HAADF-STEM image from [1 0 1] zone axis in figure 5b1 are both pure Ge ones with greater density of Ge in the brighter ones, while the Al columns are not resolvable in HAADF because of large atomic number difference between Al and Ge, but could be observed in iDPC mode in which low-dose STEM technique with four segmented on-axis annular dark field (ADF) detectors enables imaging of all elements together. Considering the lattice parameters obtained from SAED experiments and the motif coordinates determined from the projected zone axes HR-STEM images, a unit cell has been constructed using VESTA software. Projected atomic structures of the model unit cell from the aforementioned low index zones placed beside the corresponding images in figure 5(a3-c3) follow the similar distribution as in the experimental micrographs. The unit cell is shared by 24 Al atoms and 20 Ge atoms, rendering a stoichiometry of the intermetallic as $Al_6Ge_5$.

The same characterization pathway was followed to index the $β_2$ phase. Series of SAED patterns plotted commensurate with the relative tilting angles has been mapped in figure 6a. Among several low index reciprocal planar vectors (g vectors), g1, g2, and g4 were determined as principal reciprocal lattice vectors used to develop reciprocal space lattice in figure 6b. Converting them to real space lattice yields a monoclinic structure and analyzing the low index CBED patterns suggest a possible P $2_1$ space group. Lattice parameters of the monoclinic (P $2_1$) structure are given in table 4. While moving across the g3 vector along the green colored Kikuchi band, splitting of the planar peaks except g3 direction can be observed, which is an indication of twinning. This implies that the plane corresponding to the g3 vector is the possible twin plane and the orthogonal direction (into the plane) is the twin boundary direction, which has been further verified from the HR-STEM characterization (shown later). Similar HR-STEM approach like before has been employed (figure 7) to find out the atomic positional coordinates presented in table 5 and create the unit cell. Needless to mention that the atomic distributions from three low index zone axes ([001], [101], and [100]) are perfectly aligned with the respective projected atomic structures of the model unit cell constructed in VESTA software. The unit cell is equally shared by 4 atoms of each Al and Ge, yielding an intermetallic stoichiometry of AlGe.

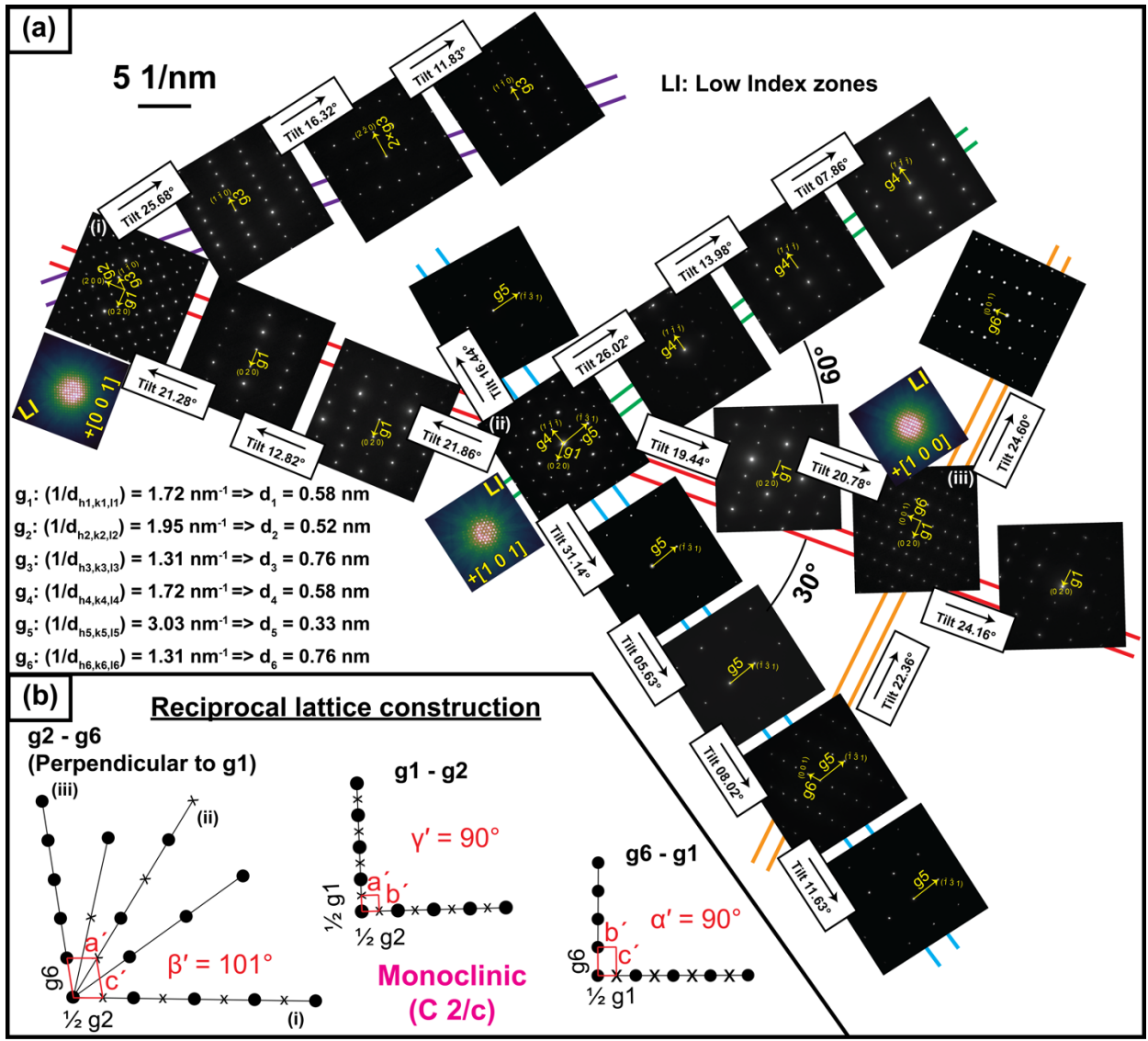

Figure 4: (a) Series of indexed SAED patterns taken from different zone axes of the $\beta_1$ phase to construct a portion of the Kikuchi pole. Three low index zone axes were determined as (i) [0 0 1], (ii) [1 0 1], and (iii) [1 0 0], the corresponding CBED patterns are placed adjacent to the diffraction patterns. (b) Relative tilting angles between the zones helped reconstruct the reciprocal space lattice with principal lattice vectors g1 ($\approx 2b'$), g2 ($\approx 2a'$), and g6 ($\approx c'$) and angles $\alpha'$ (90°), $\beta'$ (101°), and $\gamma'$ (90°). The derived lattice parameters and planar peak distribution (in terms of allowed and forbidden peaks) renders a monoclinic (C 2/c) lattice.

Table 2: Lattice parameters of the monoclinic (C 2/c) $\beta_1$ structure

| Lattice parameter | $a$ (Å) | $b$ (Å) | $c$ (Å) | $\alpha$ | $\beta$ | $\gamma$ |
|---|---|---|---|---|---|---|
| Value | 10.4 | 11.7 | 7.6 | 90° | 101° | 90° |

Table 3: <u>Atomic positions in the monoclinic (C 2/c) $\beta_1$ unit cell</u>

| No. | Atom | Positional coordinates (in fraction) | | |
|---|---|---|---|---|
| | | x (along a) | y (along b) | z (along c) |
| # 1 | Ge | -0.40 | 0.30 | -0.15 |
| # 2 | Ge | 0.00 | -0.10 | 0.25 |
| # 3 | Ge | -0.20 | 0.45 | -0.40 |
| # 4 | Al | -0.30 | 0.30 | 0.35 |
| # 5 | Al | 0.40 | 0.45 | -0.50 |
| # 6 | Al | 0.10 | 0.40 | -0.40 |

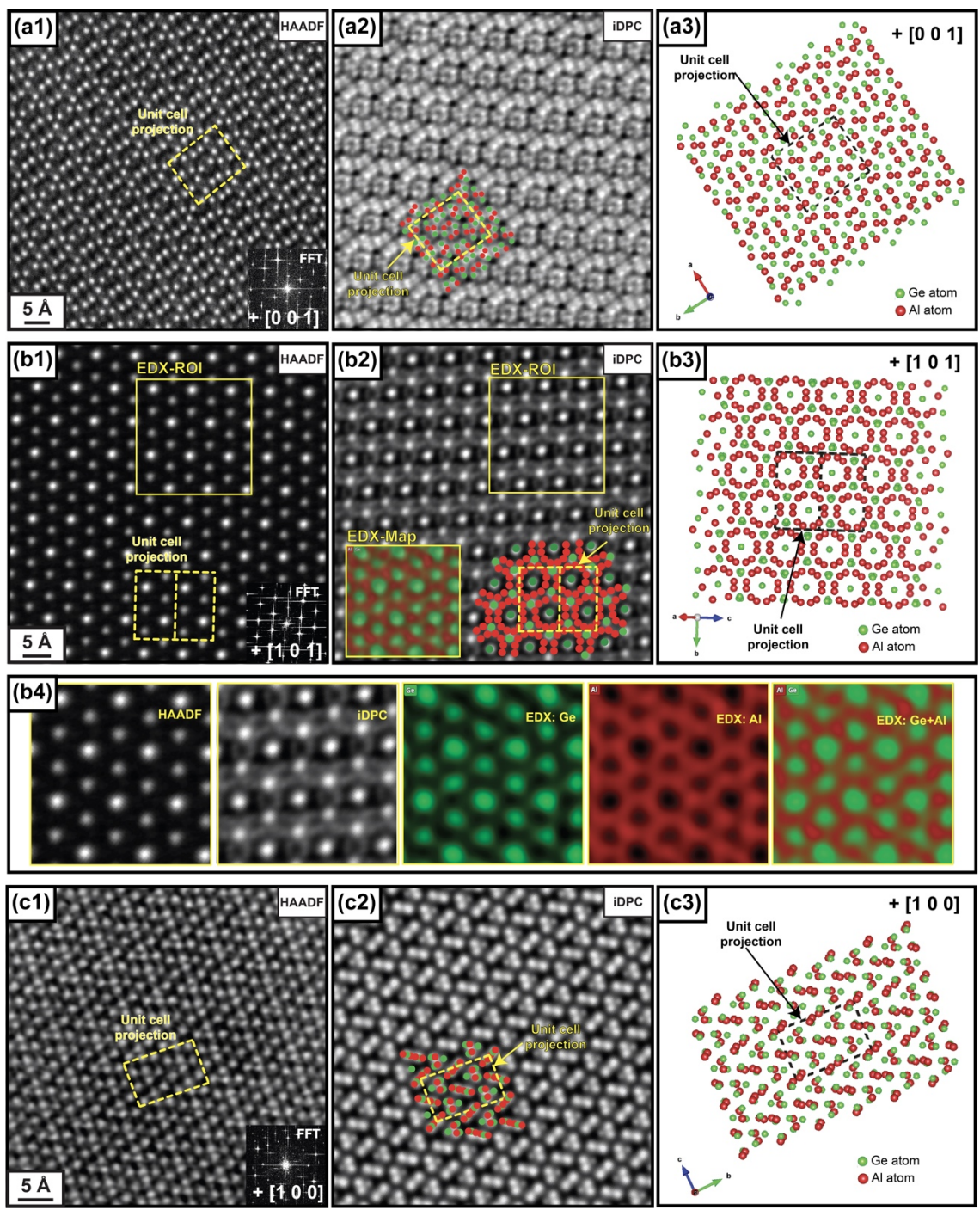

Figure 5: HAADF-STEM images of the $\beta_1$ phase along with inset FFT patterns from (a1) [0 0 1], (b1) [1 0 1], and (c1) [1 0 0] zone axes are chiefly useful to locate the Ge atomic columns because of brighter contrast arising from higher effective atomic number. Simultaneous iDPC-STEM micrographs from (a2) [0 0 1], (b2) [1 0 1] with atomic scale EDX map overlapped in inset, and (c2) [1 0 0] zone axes help pointing out the Al atomic columns as well. (b4) HAADF-STEM, iDPC-STEM, and EDX-obtained elemental maps are placed together for better understanding the nature of the atomic columns from [1 0 1] zone. Model unit cell was built up, the projection of which are given from (a3) [0 0 1] zone, (b3) [1 0 1] zone, and (c3) [1 0 0] zone. These projections are quite consistent with atomic distributions found in experimental micrographs.

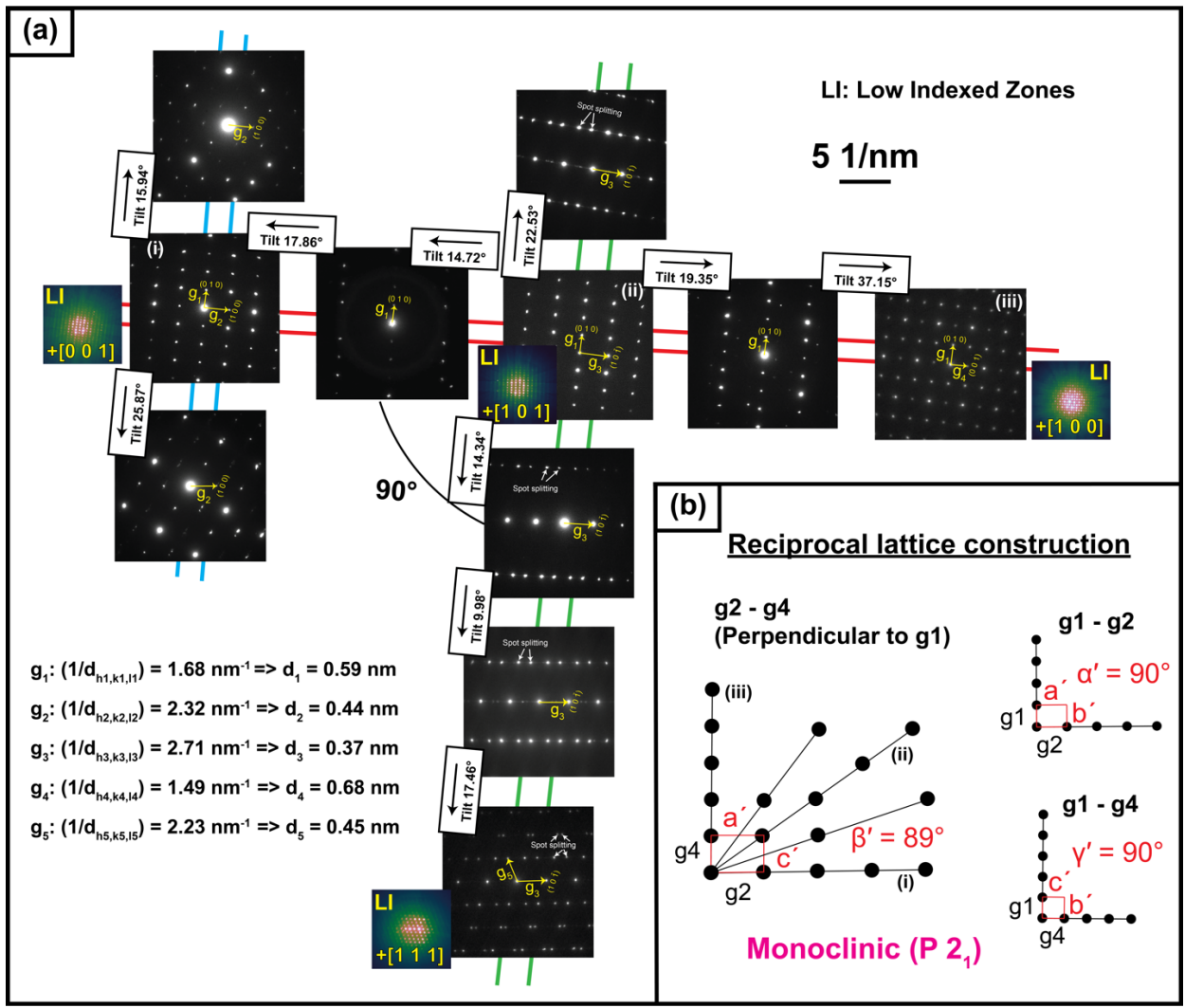

Figure 6: (a) Series of indexed SAED patterns taken from different zone axes of the $\beta_2$ phase to construct a portion of the Kikuchi pole. Three low index zone axes were determined as (i) [0 0 1], (ii) [1 0 1], and (iii) [1 0 0], the corresponding CBED patterns are placed adjacent to the diffraction patterns. (b) Relative tilting angles between the zones helped reconstruct the reciprocal space lattice with principal lattice vectors g1 ($\approx$b'), g2 ($\approx$a'), and g4 ($\approx$c') and angles α' (90°), β' (89°), and γ' (90°). The derived lattice parameters and planar peak distributions (in terms of allowed and forbidden peaks) in low index CBED patterns render a monoclinic (P $2_1$) lattice.

Table 4: Lattice parameters of the monoclinic (P $2_1$) $\beta_2$ structure

| Lattice parameter | $a$ (Å) | $b$ (Å) | $c$ (Å) | $\alpha$ | $\beta$ | $\gamma$ |
|---|---|---|---|---|---|---|
| Value | 4.4 | 5.9 | 6.8 | 90° | 89° | 90° |

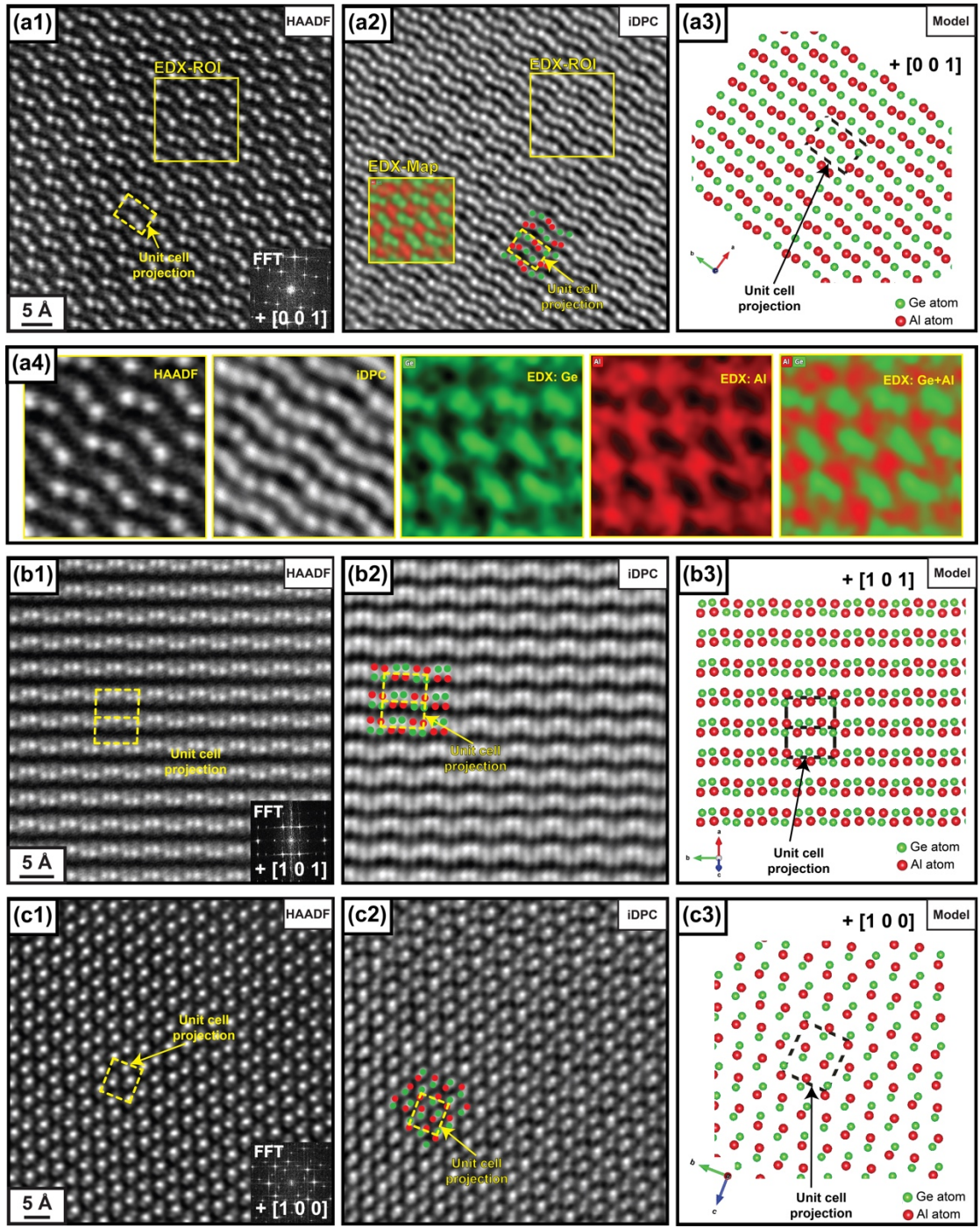

Figure 7: HAADF-STEM images of the β$_2$ phase along with inset FFT patterns from (a1) [0 0 1], (b1) [1 0 1], and (c1) [1 0 0] zone axes are chiefly useful to locate the Ge atomic columns because of brighter contrast arising from higher effective atomic number. Simultaneous iDPC-STEM micrographs from the (a2) [0 0 1] zone overlapped with atomic scale EDX map, (b2) [1 0 1] zone, and (c2) [1 0 0] zone help pointing out the Al atomic columns as well. (a4) Side-by-side HAADF-STEM, iDPC-STEM, and elemental maps help visualize the atomic columns along [1 0 1] zone. Model unit cell was built up, the projection of which are given from (a3) [0 0 1] zone, (b3) [1 0 1] zone, and (c3) [1 0 0] zone. These projections are quite consistent with atomic distributions found in experimental micrographs.

Table 5: <u>Atomic positions in the monoclinic (P $2_1$) $\beta_2$ unit cell</u>

| No. | Atom | Positional coordinates | | |
|---|---|---|---|---|
| | | x (along a) | y (along b) | z (along c) |
| #1 | Ge | 0.20 | 0.25 | 0.40 |
| #2 | Ge | -0.30 | 0.05 | -0.10 |
| #3 | Al | 0.20 | 0.00 | 0.05 |
| #4 | Al | -0.30 | 0.30 | -0.45 |

Based on the experimental information about the lattice structure and motifs, unit cells of $\beta_1$ and $\beta_2$ intermetallics were constructed. The corresponding CIFs (after DFT relaxation) could be found in supplementary information. For future reference, the interplanar spacing values corresponding to the low index planes are reported in table 6, where the Miller indices (h k l) of the planes have been derived from the SAED experiments and also verified by fitting the values to the equation used for determining interplanar spacing in monoclinic systems:

$$\frac{1}{d^2} = \frac{1}{sin^2\beta} \times \left[\frac{h^2}{a^2} + \frac{k^2 sin^2\beta}{b^2} + \frac{l^2}{c^2} - \frac{2hl cos\beta}{ac}\right] \tag{1}$$

Table 6: <u>List of low indices interplanar spacings (d) of the $Al_6Ge_5$ and AlGe intermetallics</u>

| Intermetallic: $Al_6Ge_5$ | | | Intermetallic: AlGe | | |
|---|---|---|---|---|---|
| $d_i$ marked in figure 4 | $d_i$ values (nm) | Miller indices of the planes (h k l) | $d_i$ marked in figure 6 | $d_i$ values (nm) | Miller indices of the planes (h k l) |
| $d_1$ | 0.581 | (0 2 0) | $d_1$ | 0.592 | (0 1 0) |
| $d_2$ | 0.517 | (2 0 0) | $d_2$ | 0.435 | (1 0 0) |
| $d_3$ | 0.763 | (1 -1 0) | $d_3$ | 0.369 | (1 0 -1) |
| $d_4$ | 0.581 | (1 -1 1) | $d_4$ | 0.677 | (0 0 1) |
| $d_6$ | 0.761 | (0 0 1) | $d_5$ | 0.448 | (-1 1 0) |

### 3.3. <u>Other forms of metastability: Defect structures and precipitation</u>

An important yet often overlooked feature of metastability in the rapid solidified microstructures is high density of defects, such as dislocations and stacking faults, because of solidification shrinkage that leads to stress induced plastic flow in the forming solids [36, 37]. The atomic-scale images earlier were intended for crystallographic indexing of the phases; hence they represent defect-free regions only. However, all the phases within the melt pool exhibit high density of solidification induced defects. Formation of dislocation lines and loops have been noted in Al-rich $\alpha$ phases from both zone 1 and zone 2 within the melt pool, as appear in the in-zone bright field (BF) STEM images in figure 8a and 8d. One from each of these high dislocation density regions have been explored in figure 8 b1 and e1 using HR-STEM from the same [0 1 1] zone axes. One set of (1 1 1) planes have been masked from the FFT patterns to reconstruct the IFFT images in figure 8 b2 and e2, which evince edge dislocations present in these planes. Interestingly, some evidence of formation of stacking faults and twins have been recorded in the {1 1 1} planes along <1 1 2> directions in the α phases, shown in figure 8c and 8f. In order for a stacking fault or twin to nucleate in an FCC material, the system has to meet a certain free energy requirement that can be expressed as [38]:

$$\Delta G_v > \gamma_{sf}/h \tag{2}$$

which can be further simplified to:

$$\frac{\Delta H_m \times \Delta T}{\Omega \times T_m} > \frac{\gamma_{sf}}{h} \tag{3}$$

where $\Delta G_v$ is volume free energy, $\Delta H_m$ is enthalpy of fusion, $\Omega$ is molar volume, $\Delta T$ is undercooling, $T_m$ is melting point, $\gamma_{sf}$ is stacking fault energy along primary slip planes {111}, and $h$ is interplanar spacing of {111} planes. It is therefore obvious that higher degree of undercooling is more favorable for twinning, which is why the planar faults may have formed during rapid solidification.

On the other hand, the intermetallic rich phases are not exceptions as far as solidification instigated defects are concerned. As revealed by the annular dark field (ADF) STEM images, $\beta_1$ phase exhibits dislocation network connecting the α phases (figure 8g1), the density of which is more prominent near the outskirts of zone 1 where the α lamellae are decomposed (figure 8g2). Defect structure in the $\beta_2$ phase is manifested by high density of nano-twins displayed by the BF-STEM micrograph in figure 8h1. The corresponding HAADF-STEM image is kept at the inset to help locate the $\beta_2$ phases. Figure 8h2 is an ADF HR-STEM image of the nano-twins from [2 -1 2] zone axis, according to which the twin plane has been identified as {1 0 -1} and the twin boundary direction should be <1 4 1>.

Another form of metastability is precipitation in the α phases both in zones 1 and 2, the extent of which is more in the former. Locally precipitate rich regions are highlighted by HAADF STEM micrographs from the α plates of zone 1 and zone 2 in figure 9a and figure 9c, respectively. The diameters of these precipitates are in the range of 2-5 nm. HAADF and ABF HR-STEM images of the precipitates from each of the zones have been presented in figure 9(b1-b2) and figure 9(d1-d2). As evident from the [0 1 1]$_{Al}$ zone axis atomic resolution images, the precipitates have FCC lattice with similar lattice parameter as Al, making them fully coherent with the Al-rich α matrix. Brighter contrast in the HAADF STEM images in essence indicates Ge rich compositions of these clusters, which is further verified by EDX-based elemental mapping in figure 9(b3, d3). It is worth noting that these Ge-rich atomic clusters in α phases may have reduced the SFE significantly (as Ge has a lower SFE than Al), which brings about the stacking faults and twins in the α phases generated from rapid solidification induced shrinkage.

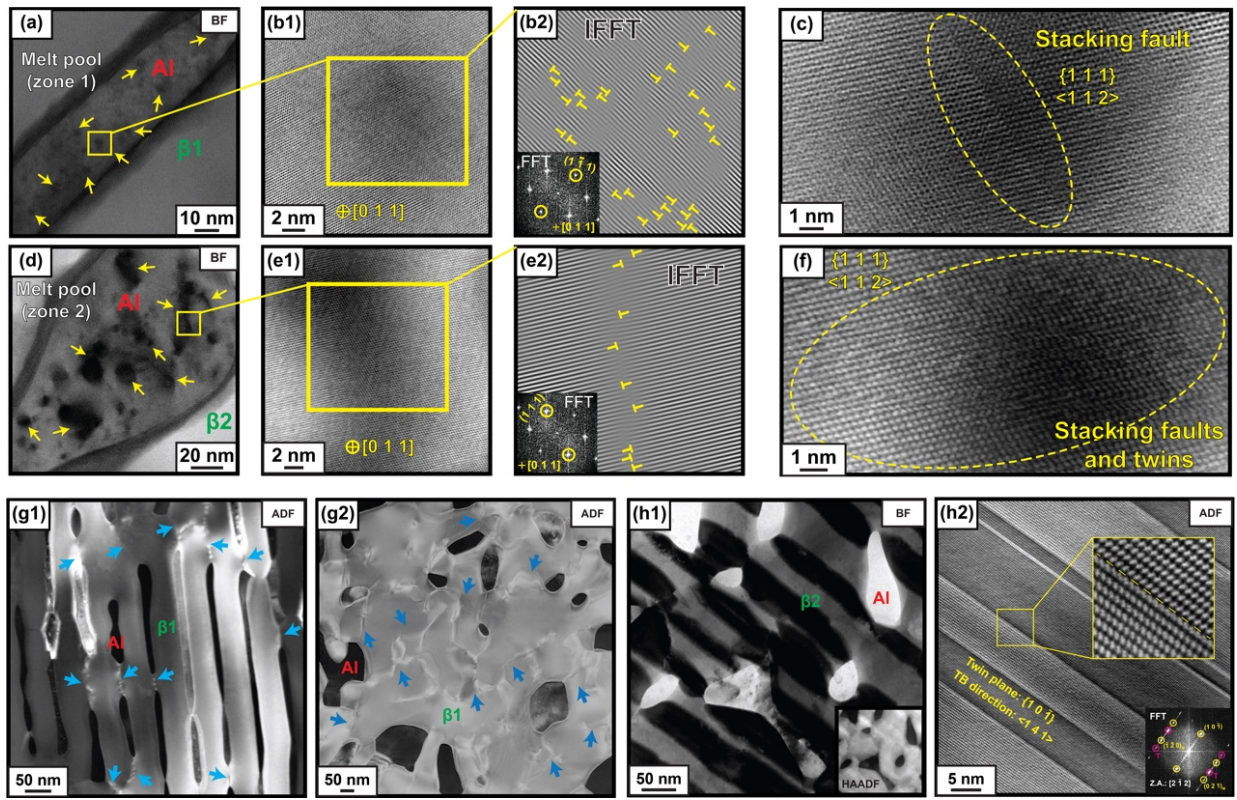

Figure 8: BF-STEM images of dislocation lines and loops, HR-STEM images of the dislocation heavy regions, IFFT reconstructions of one set of (1 1 1) planes revealing the presence of edge dislocations, and HR-STEM images pointing out planar faults, respectively, for the α phases in (a, b1, b2, c) zone 1 and (d, e1, e2, f) zone 2 of the melt pool. (g1, g2) In-zone ADF-STEM images displaying dislocation network in $\beta_1$ phase presumably in the prismatic planes as identified from the inset zone axis diffraction patterns. (h1) BF-STEM micrograph of nano-twins in the $\beta_2$ phase with the corresponding HAADF-STEM image at the inset. (h2) ADF HR-STEM image of the nano-twins with corresponding FFT pattern at the inset that points out {1 0 -1} as the twin plane and <1 4 1> as the twin direction.

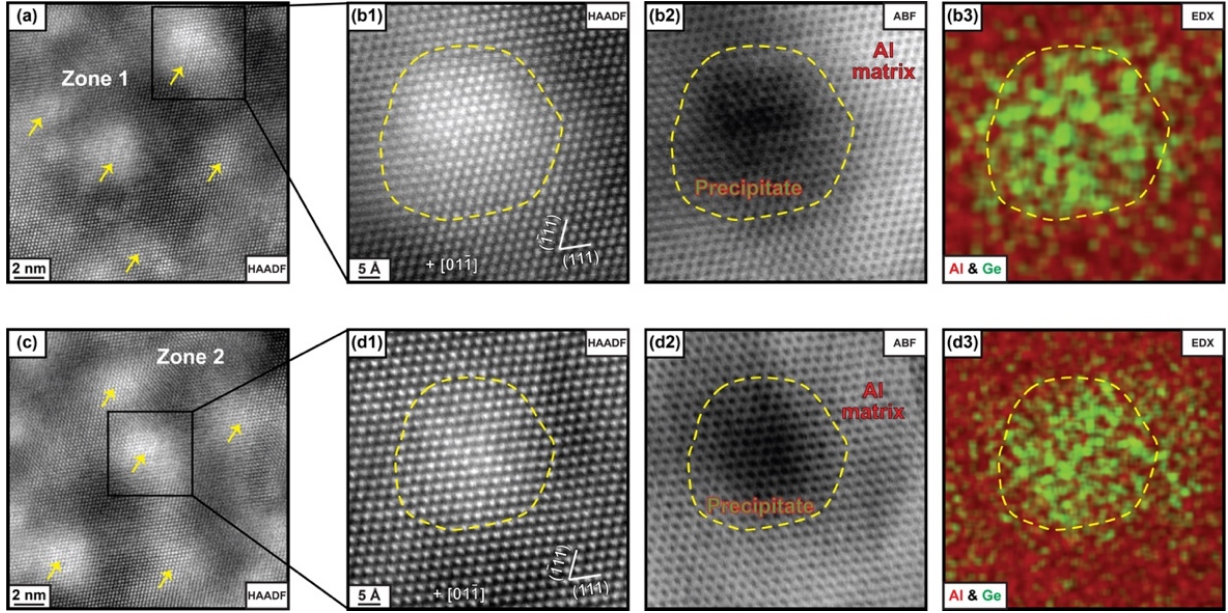

Figure 9: HAADF STEM images of locally precipitate dense regions of the α phases in (a) zone 1 and (c) zone 2. Simultaneous HAADF STEM, ABF STEM, and EDX generated elemental maps of the Ge rich precipitates respectively from (b1, b2, b3) zone 1 and (d1, d2, d3) zone 2 within the melt pool.

## 4. Discussion

The experimental findings have depicted formation of solid solution (Al-rich) – intermetallic (Ge-Al) two phase microstructures having plate-like morphologies as a result of rapid solidification of Al-Ge hypereutectic cast alloy. The complex crystal structures of the intermetallic-rich phases have been identified as monoclinic (C 2/c) and monoclinic (P $2_1$) using SAED and HR-STEM, which entails further validation. Also, it is essential to have fundamental understanding of the reasons for evolution of different metastable phases and morphologies at nano-scale across the melt pool. This section aims to address all the issues pointed above in detail.

### 4.1. Structural validation of the metastable phases by DFT calculation and HR-STEM image simulation

Formation of metastable crystalline and amorphous phases have been reported before in Al-Ge system in response to non-equilibrium processing such as melt spinning, splat quenching, thin film deposition, or mechanical alloying [25]. Table 7 enlists the reported metastable crystalline phases along with their crystallography and alloy composition.

Table 7: Metastable crystalline phases reported in Al-Ge system

| Starting composition (at. %) | Crystal structure <Space group> | Lattice parameters (lengths in *nm*, angles in °) | Reference |
|---|---|---|---|
| Al – 10 Ge | Tetragonal <I4/mcm> | $a = 0.613$, $c = 0.492$ | [39] |
| Al – 33 Ge | Cubic | $a = 1.287$ | [40] |

| Al – 33 Ge | Tetragonal | $a = 1.291, c = 1.200$ | [41] |
| Al – 37.5 Ge | Tetragonal | $a = 0.659, c = 1.201$ | [42] |
| Al – 45 Ge | Rhombohedral <R3C> | $a = 0.767; \beta = 96.55$ | [25, 43] |
| Al – (42-45) Ge | Hexagonal <P6/mmm> | $a = 1.40, c = 0.72$ | [24] |
| Al – (42-45) Ge | Hexagonal | $a = 1.35, c = 0.71$ | [44] |
| Al – (41-50) Ge | Orthorhombic <P b c a> | $a = 0.78, b = 0.57, c = 0.73$ | [24] |
| Al – 50 Ge | Cubic | $a = 1.381$ | [41] |
| Al – 50 Ge | Monoclinic <P2$_1$/c> | $a = 0.673, b = 0.582, c = 0.805; \beta = 147.85$ | [24, 43] |

According to Table 7, changes in non-equilibrium process parameters result in different phases even for identical alloy compositions. The same is true for compositional variation, i.e., a modest change in alloy chemistry produces different metastable phases for similar processing conditions. Similar compositions as this study (Al – 32 at.% Ge) were indeed used before, but not the same processing condition. Fine-spot laser surface remelting reportedly imparts cooling rate of as high as $10^7$ K/s in Al alloys [16, 28], which is considerably higher than the cooling rates for non-equilibrium solidification in previously reported specimens, such as in melt spun ribbons. Detailed crystallographic analyses have elucidated that $\beta_1$ and $\beta_2$ are the intermetallic rich metastable phases that have not been reported before, which can be attributed to the unorthodox processing route. Even though there is a reported monoclinic structure in this system, the alloy composition as well as the lattice parameters differ remarkably from the monoclinic structures found in this study.

CIFs of the intermetallics modeled in VESTA were used for DFT-based geometric optimization using VASP. It should be noted that the optimized, i.e., relaxed structures have nearly same lattice parameters as they have been predicted from the experiments. The purpose of DFT optimization is therefore to predict the structural stability and to fine tune the lattice parameters as well as motif coordinates. The relaxed unit cells of Al$_6$Ge$_5$ and AlGe are portrayed in figure 10 (a, b). Corresponding interactive CIFs could be found in supplementary information. Total energies ($E_{tot}$) of the two structures were computed to be -177.7958 eV and -32.6197 eV, respectively. The formation energies per atom ($E_{form}$) of these Al$_x$Ge$_y$ structures have been calculated as [45]:

$$E_{form} = \frac{E_{tot} - n_{Al}E_{Al} - n_{Ge}E_{Ge}}{n_{Al} + n_{Ge}} \quad (4)$$

where $n_{Al}$ and $n_{Ge}$ are the number of Al and Ge atoms in the unit cells, and $E_{Al}$ and $E_{Ge}$ are the solid state energies of Al and Ge, respectively. Equation 4 yields $E_{form}$ of the monoclinic (C 2/c) Al$_6$Ge$_5$ and the monoclinic (P 2$_1$) AlGe unit cells to be 0.04373 eV/atom and 0.04253 eV/atom, respectively. Formation energies slightly above zero indicate the influence of kinetic precursors, i.e., undercooling, and cooling rate, on the production of these non-equilibrium intermetallics. Further verification of these structures has been done by simulating HR-STEM images using three different detectors (HAADF, ABF, and iDPC) from specific low index zone axes ([101] zone for β₁ and [100] zone for β₂] in figure 10 which perfectly match the experimentally obtained HR-STEM images.

The experimentally obtained structures were refined during the DFT optimization. The refined lattice parameters and motif coordinates reported in tables 2, 3, 4, and 5 have been reported in tables 8, 9, 10, and 11 below.

Table 8: Refined lattice parameters of the monoclinic (C 2/c) β₁ structure after DFT relaxation

| Lattice parameter | $a$ (Å) | $b$ (Å) | $c$ (Å) | $\alpha$ | $\beta$ | $\gamma$ |
|---|---|---|---|---|---|---|
| Value | 10.32033 | 11.56400 | 7.65800 | 90° | 100.4557° | 90° |

Table 9: Refined atomic positions in the monoclinic (C 2/c) β₁ unit cell after DFT relaxation

| No. | Atom | Positional coordinates (in fraction) | | |
|---|---|---|---|---|
| | | x (along a) | y (along b) | z (along c) |
| # 1 | Ge | -0.392250 | 0.250980 | -0.142530 |
| # 2 | Ge | 0.000000 | -0.102450 | 0.250000 |
| # 3 | Ge | -0.175230 | 0.425600 | -0.400210 |
| # 4 | Al | -0.278810 | 0.292960 | 0.327990 |
| # 5 | Al | 0.379200 | 0.451790 | -0.484050 |
| # 6 | Al | 0.076780 | 0.406860 | -0.428760 |

Table 10: Refined lattice parameters of the monoclinic (P 2₁) β₂ structure after DFT relaxation

| Lattice parameter | $a$ (Å) | $b$ (Å) | $c$ (Å) | $\alpha$ | $\beta$ | $\gamma$ |
|---|---|---|---|---|---|---|
| Value | 4.35400 | 5.92200 | 6.77415 | 90° | 89.9105° | 90° |

Table 11: Refined atomic positions in the monoclinic (P 2₁) β₂ unit cell after DFT relaxation

| No. | Atom | Positional coordinates | | |
|---|---|---|---|---|
| | | x (along a) | y (along b) | z (along c) |
| #1 | Ge | 0.266690 | 0.237250 | 0.389490 |
| #2 | Ge | -0.283080 | 0.030020 | -0.110190 |
| #3 | Al | 0.217600 | -0.007600 | 0.071130 |
| #4 | Al | -0.234560 | 0.274320 | -0.429320 |

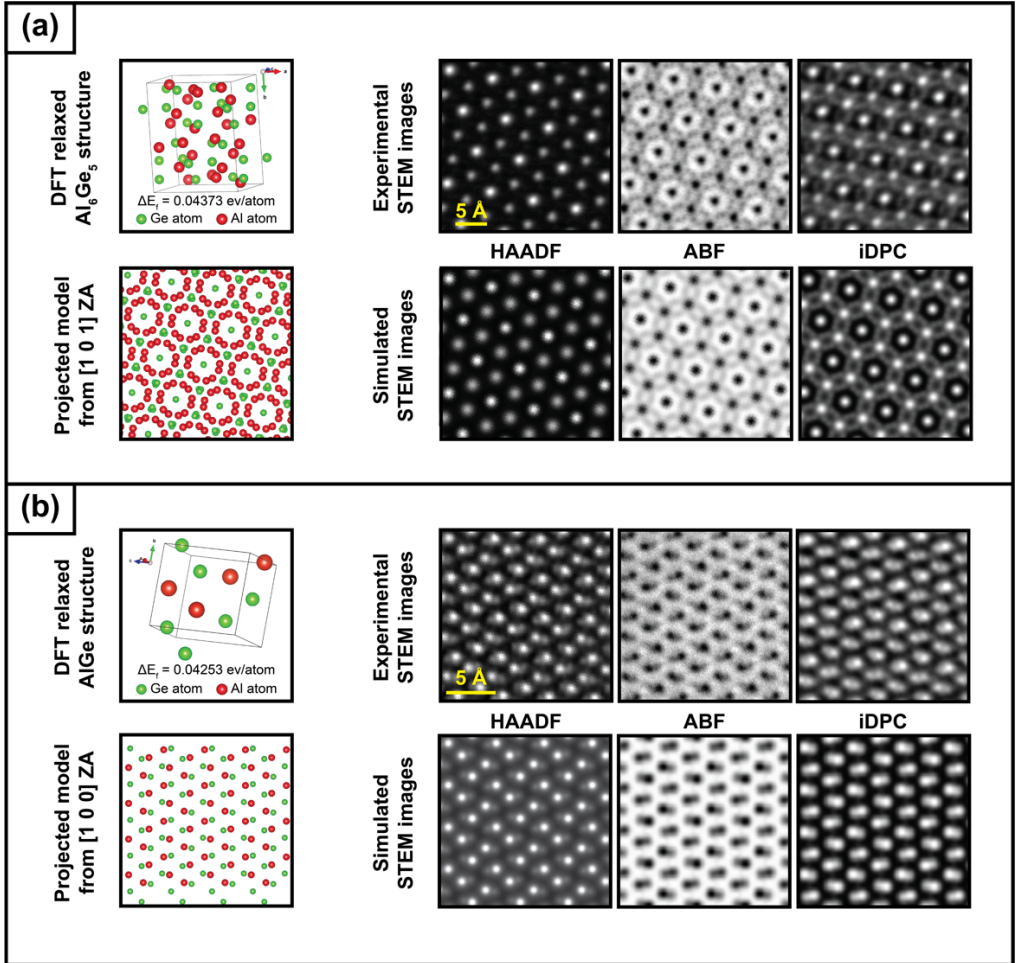

Figure 10: DFT relaxed unit cell and projected atomic structure model from (a) [1 0 1] zone axis of $Al_6Ge_5$ and (b) [1 0 0] zone axis of AlGe intermetallics that were used to simulate atomic resolution HAADF, ABF, and iDPC STEM images which perfectly match with the experimental micrographs.

Apart from the energy calculation based validations, another degree of verification could be done by comparing the stoichiometry of the intermetallics with the elemental quantification given in table 1. To be specific, $Al_6Ge_5$ and AlGe intermetallics should contain approximately 45.4 at. % and 50 at. % of Ge, respectively, which are quite similar to the EDX obtained data.

### 4.2. Metastable phase equilibria and correlation with microstructure

As observable from the as-remelted microstructure, the metastable phases generate metastable phase equilibria with α only, and the similar equilibria has not been noted anywhere else (to our knowledge). Thus, from the viewpoint of thermodynamics, it is reasonable to argue that in both of the two microstructural zones within the melt pool, the metastable invariant reactions should be expressed as Liquid → α + metastable phase ($β_1$ or $β_2$), but not as Liquid + $β_1$ (or $β_2$) → $β_2$ (or $β_1$). On that account, even under non-equilibrium conditions, the invariant reactions here is a 'eutectic' rather than 'peritectic'. In support of this point, the two-phase microstructures of both zones are consistent with eutectic-like patterns reported elsewhere [46, 47].

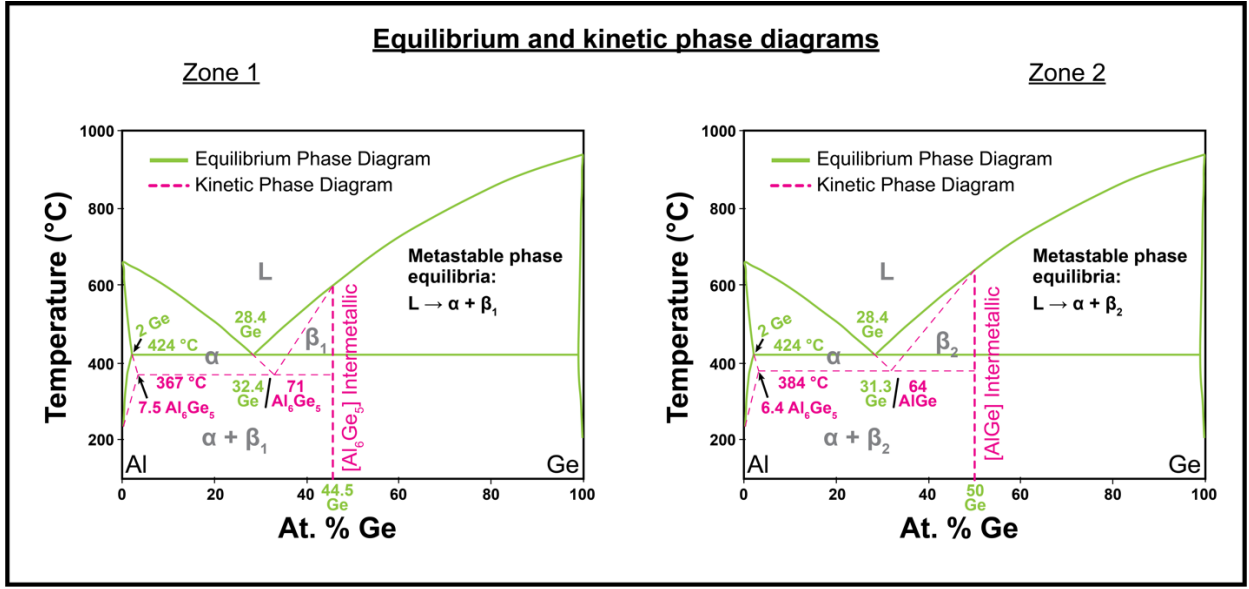

Figure 11: Kinetic phase diagrams for each of the metastable eutectic reactions in zone 1 and zone 2 overlapped on the equilibrium Al-Ge phase diagram.

Kinetic phase diagrams corresponding to these metastable eutectic reactions have been drawn and compared to the equilibrium eutectic phase diagram in figure 11. As the melt pool microstructure is fully eutectic, without the presence of any primary Al or Ge or other metastable dendritic phases, the alloy compositions in different zones (Al – 32.4 at.% Ge in zone 1 and Al – 31.3 at.% Ge in zone 2) can be treated as the invariant points for the metastable eutectic reactions between Al and the experimentally determined possible intermetallics, namely, $Al_6Ge_5$ and AlGe. The slight difference in composition of the eutectic points results in markedly distinct eutectic temperatures for the two eutectic systems due to the pronounced slope of the α liquidus (−8.49 °C/at%). That is, by extrapolating the α liquidus to the metastable eutectic compositions, we obtain approximately the eutectic temperatures of 384 °C and 367 °C. As far as the metastable phase diagrams are concerned, the solid solubility limit of Ge in α would be higher (7.5 at.% and 6.4 at.%, respectively) than the equilibrium solid solubility of Ge in α, which is ~2 at. % at 424°C. These increased solid solubilities imply a supersaturation during subsequent cooling that then led to solid state phase transformation [21], i.e., formation of Ge-rich precipitates within the α phases. The precipitation occurred at atomic scale, which helps them maintaining coherency with α [48].

Based on the preceding analysis, in the metastable phase diagrams of Al – $Al_6Ge_5$ and Al – AlGe, the eutectic points have been estimated as Al – 71 at.% $Al_6Ge_5$ and Al – 64 at.% AlGe. These compositions can be converted to wt.% using the atomic weights of the elements and rewritten as approximately Al – 68 wt.% $Al_6Ge_5$ and Al – 72 wt.% AlGe, respectively. Therefore, the metastable phase diagrams in terms of mass fraction are more or less symmetric about the eutectic points. As a rule of thumb, symmetric phase diagrams usually correspond to near-equal volume fraction of lamellar or degenerated eutectic phases [49]. Precise calculation of volume fraction of the phases from their weight ratio is non-trivial in this case since the densities of the new phases are still unknown; but it is still possible to envisage the apparent volume fraction of the phases in the $α_1$-$β_1$ and $α_2$-$β_2$ microstructures from figure 1(c-d). In that case, an assumption should be made that the microstructures are uniform along all three dimensions, which can be justified by the fact that cross-section STEM micrographs present similar length scale and morphological distribution as the SEM images from the surfaces. Therefore, the volume fraction will be close to the area fraction of the phases present, which can be approximated from the phase plate thickness ratio for lamellar microstructures. The estimation yields volume fractions of metastable phases in zone 1 and zone 2 to be around 66% and 50%, respectively. These approximate values fall above the classical (thermodynamic) threshold of volume fraction of the second phase ($f_v \geq 0.314$), which is a condition for plate-like morphology in eutectics [50]. These plate-like structures are mostly regular throughout and in order to quantitatively determine the reasons for absence of irregularities like facets, other important parameters (e.g., entropy of fusion) should be considered, but those are very difficult to estimate for these newly found metastable phases.

The microstructures of the two eutectics are different, yet related. For instance, the degenerated nature of the eutectic in zone 2 may arise from fragmentation of the α plates in zone 1. In fact, the fragmentation of the plates while moving toward the melt pool exterior (zone 2) from the melt pool core (zone 1) have been recorded in STEM and shown in figure 12 (a-c), which are consistent with the STEM tomography reconstructions as well (see supplementary video 1 and 2). As expected for uniform plate-like growth of two eutectic phases, there exists a prominent orientation relationship (OR) between α and $β_1$. This well-defined and consistent growth characteristic is reflected in the HR-STEM image of the α-$β_1$ interface in figure 12d. Comparison of the FFT patterns obtained from the FCC α and monoclinic $β_1$ establishes the following OR:

$(2\ 0\ 0)_α \parallel (0\ -1\ 0)_{β_1}$; $[0\ 1\ 1]_α \parallel [2\ 0\ 1]_{β_1}$

The fragmented α plates could not maintain any well-defined growth crystallography with respect to the $β_2$ phase. Figure 12e is the HR-STEM micrograph of an α-$β_2$ interface, where the α phase is in the same zone [0 1 1] as figure 12d, which is evident from the corresponding FFT pattern. However, the FFT pattern from $β_2$ makes it clear that the same phase is not even close to a zone axis. The same observation was noted for other zone axes of α (like [0 0 1] and [1 1 1]). Therefore, there does not exist any prominent OR between these two phases in zone 2.

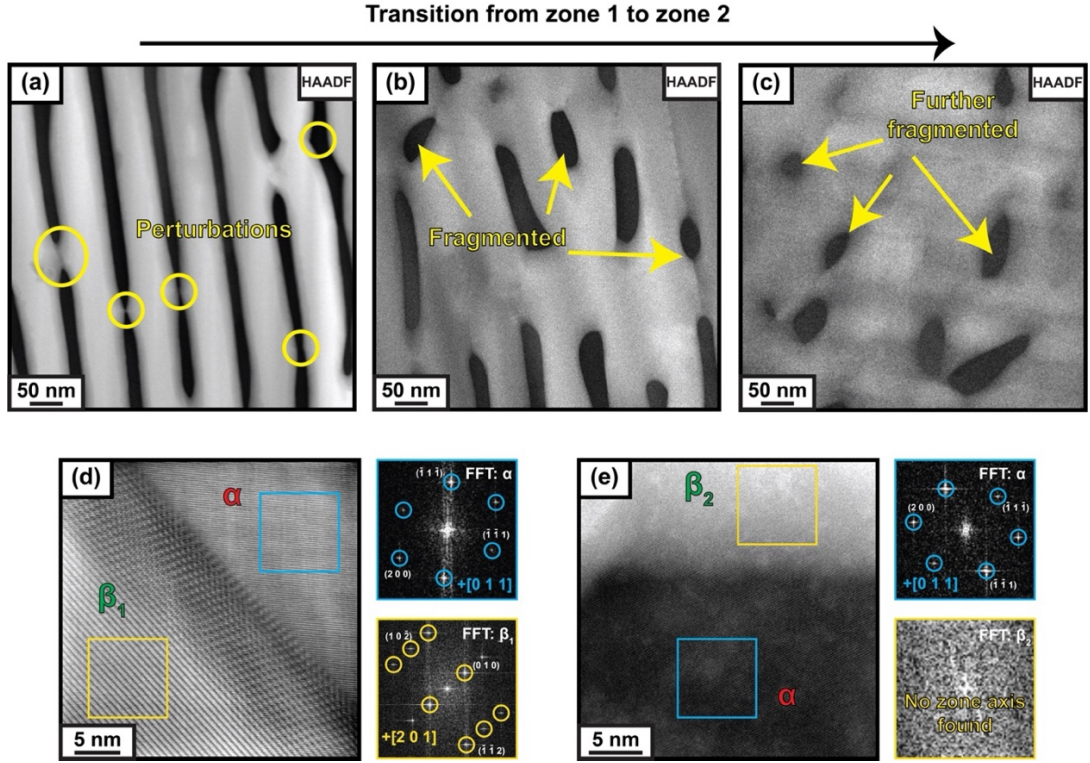

Figure 12: HAADF STEM micrographs unveiling (a) uniform distribution of α plates with some perturbations in the core region of zone 1, (b) fragmented α plates at the outskirts of zone 1, and (c) even more fragmented α plates exhibiting particle like distribution in zone 2. Dark contrast is α phase. ADF HR-STEM images of the interphase interfaces along with FFT patterns from respective phases showing (d) strong orientation relationship between α and $β_1$, (e) while no significant orientation relationship could be established between α and $β_2$.

Apart from plate-like morphologies and unusual phase evolution, nano-scale spacing of these microstructures is another interesting attribute and must be the direct consequence of the high cooling rates associated with rapid solidification. Fine-spot laser (~ 75 μm spot diameter) induces higher linear laser energy density and as a result, greater cooling rate [16, 51], which provides greater driving force (undercooling) for microstructural refinement. The refinement is however opposed by the capillary effect by setting a theoretical limit of minimum achievable interlamellar spacing during solidification [52]. Microstructural refinement using laser melt quenching is also limited by the formation of keyhole porosities at very high cooling rates.

### 4.3. Competitive growth of stable and metastable eutectics

The above discussion provides new insight on phase equilibria in the Al-Ge system, but it does not account for *why* the metastable eutectics form within the melt pool. To explain the selection of microstructure under LSR, we invoke (as a first-order approximation) the so-called competitive growth criterion or extremum criterion [53], which states that the growth form that dominates is the one with the highest interface temperature ($T^*$) for a given composition, velocity ($V$), and thermal gradient. It is assumed that that nucleation of the various phases can actually occur, i.e., it is not rate-limiting.

In figure 13, plausible $T^*$ *versus* $V$ curves have been plotted for the three growth forms under consideration here — the α–Ge, α–$β_2$, and α–$β_1$ eutectics — in order to help rationalize our results.

For all three eutectics, we let the undercooling $\Delta T \equiv T_E - T^*$ (where $T_E$ is the eutectic temperature retrieved from sec. 4.2) increase monotonically with $V$. Accordingly, at low $V$, the thermodynamically stable ɑ–Ge eutectic is to be expected. Importantly, this will *always* hold true for interface temperatures $T^*_{ɑ-Ge}$ that exceed the metastable eutectic temperatures $T_{E,ɑ-β_2}$ and $T_{E,ɑ-β_1}$. With increasing $V$, it is conceivable that the undercooling of the ɑ–Ge eutectic increases at a faster rate than that of the ɑ–β₂ eutectic, which would bring about a transition from stable to metastable eutectic. At even higher velocity, a competition between the two metastable eutectics is anticipated: eventually, ɑ–β₁ will possess a higher interface temperature than ɑ–β₂ and will take over the microstructure. The higher velocity implies a more refined microstructure, which is why the average interlamellar spacing is generally smaller in zone 1 compared to zone 2. A similar line of reasoning is used in [54, 55] to explain the solidification conditions that favor the formation of white *versus* gray cast Fe-C eutectics, and in [56] for the formation of irregular *versus* spiral Zn-Mg eutectics, albeit at far lower velocity and undercooling in both cases.

The schematic diagram in figure 13 indicates that microstructure selection in Al-Ge is highly sensitive to the *local* interfacial velocity. This result can be interpreted in the context of LSR: in moving from bottom to top of the melt pool, the velocity will increase rapidly and eventually approach the (imposed) laser-beam velocity (see inset); ultimately, the local $V$ can vary by three to four orders-of-magnitude as a function of melt-depth [57]. In contrast, we find that the variation in composition across the melt pool is relatively weak (within ~1 at.% based on our STEM-EDX data), which would indicate that microstructure selection here is likely not the result of a constitutional effect but rather of the local solidification conditions.

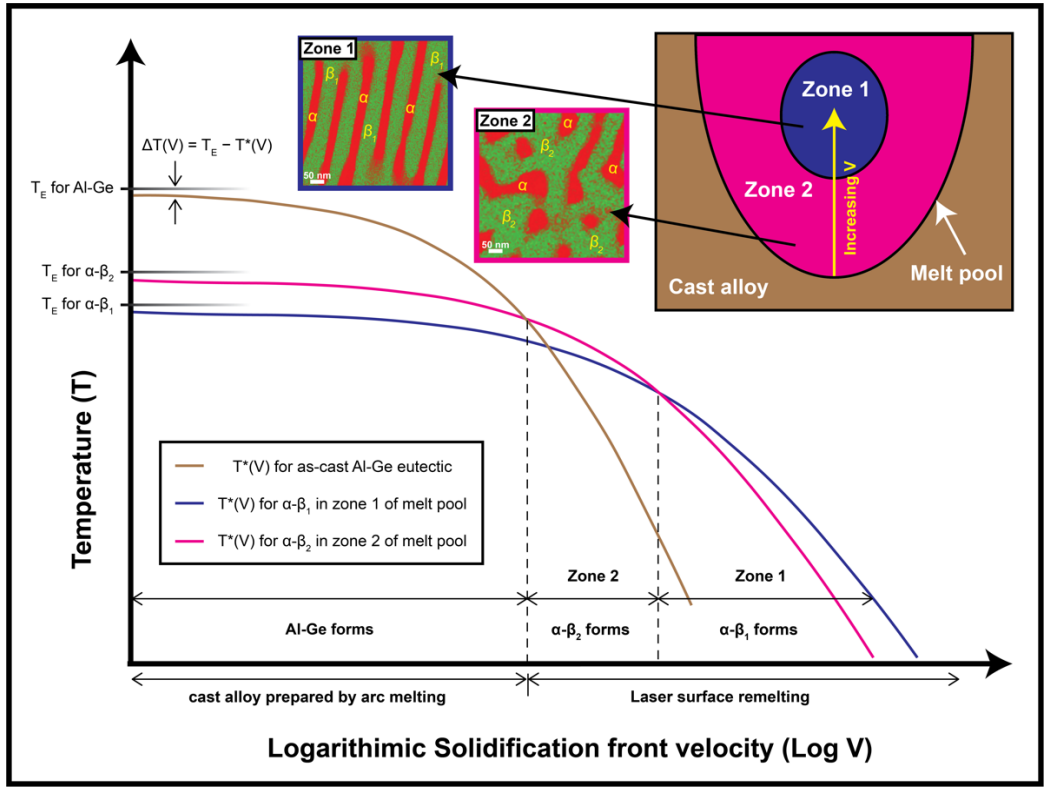

Figure 13. Schematic diagram showing the growth temperatures of three eutectics (ɑ–Ge, ɑ–β₂, and ɑ–β₁) in relation to the solidification velocity. Not drawn to scale.

## 5. Summary and conclusions

The results obtained by advanced atomic scale characterizations in this investigation show new metastable phases, with high density of dislocations and planar faults, in the Al-Ge eutectic microstructure as a result of laser rapid solidification. Key takeaways have been summarized below:

- Laser surface remelting of Al – 32 at.% Ge eutectic alloy yields ultrafine plate-like microstructures with Al-rich FCC α phase and Al-Ge intermetallics. Using a combination of TEM-SAED, STEM EDX, HR-STEM, and atomic-scale modeling, possible crystal structures of the newly found intermetallic phases have been indexed as monoclinic <C 2/c> [$Al_6Ge_5$] and monoclinic <P $2_1$> [AlGe]. The experimentally predicted structures could be relaxed by DFT and the simulated HR-STEM images perfectly match with the experimental micrographs, which essentially cross validates the indexing accuracy. Values of formation energies per atom slightly above zero imply the metastability of the intermetallic phases, i.e., in equilibrium condition, formation of these phases may not be feasible, which is why they are only present in the rapid solidified microstructures.

- Corresponding Al – $Al_6Ge_5$ and Al – AlGe metastable eutectic phase diagrams have been developed and used to correlate with plate-like morphological evolution. Decomposed lamellar structure in zone 2 has been attributed to fragmented plates of α, where there is no preferred crystallographic orientation relationship between the two phases, unlike in zone 1.

- Microstructure selection is rationalized on the basis of competitive growth between stable and metastable eutectics. At low solidification velocity, typical of casting, the thermodynamically stable Al-Ge eutectic is favored, while at the far higher velocities accessible in LSR, the metastable eutectics can emerge.

- Other forms of metastability in the rapid solidified microstructure have been manifested by defects and precipitation. Defects such as dislocations and planar faults are common in both of the Al-rich and the intermetallic-rich phases across the melt pool which primarily arise as a result of rapid solidification induced shrinkage and consequent plastic flow. Whereas, increased solid solubilities of Al in the metastable intermetallics have resulted in precipitation in the α phases in response to rapid quenching.

Future studies will focus on the evolution of different physical properties in these unusual metastable nano-scale Al – $Al_xGe_y$ microstructures and will also explore the influence of the defects as well as precipitates on various properties.

## Acknowledgements

This work is funded by DOE, Office of Science, Office of Basic Energy Sciences with the grant number of DE-SC0016808. AJS acknowledges support from DOE under grant DE-SC0023147. Arc melting of the materials were done at the Materials Preparation Center (MPC) at Ames Laboratory, Iowa, USA. Experimental characterization was performed in the Michigan Center for Materials Characterization [(MC)$^2$] at the University of Michigan - Ann Arbor. Authors acknowledge assistance of Dr. Mohsen Taheri Andani in laser remelting experiments.

## CRediT Author Statement

**Arkajit Ghosh:** Conceptualization, Experiments (LSR, SEM, TEM, STEM-EDX, HR-STEM), Validation (atomic modeling), Formal analysis, Writing - original draft. **Wenqian Wu:** Validation (DFT). **Tao Ma:** Validation (tomography, image simulation). **Ashwin J Shahani:** Conceptualization, Formal analysis, writing – review & editing. **Jian Wang:** Writing - review & editing, Project administration. **Amit Misra:** Conceptualization, Writing - review & editing, Supervision, Project administration.

## Declaration of Competing Interest

The authors declare that they have no known competing financial interests or personal relationships that could have appeared to influence the work reported in this paper.